\documentclass[a4paper]{jpconf}

\newcommand{\be}{\begin{equation}}\newcommand{\ee}{\end{equation}}
\newcommand{\bea}{\begin{eqnarray}}\newcommand{\eea}{\end{eqnarray}}
\newcommand{\beaa}{\begin{eqnarray}}\newcommand{\eeaa}{\end{eqnarray}}
\newcommand{\ba}{\begin{array}}\newcommand{\ea}{\end{array}}
\newcommand{\bit}{\begin{itemize}}\newcommand{\eit}{\end{itemize}}
\newcommand{\ben}{\begin{enumerate}}\newcommand{\een}{\end{enumerate}}

\def\1{{_{1}}}\def\2{{_{2}}}
\def\ZzZ{{\hbox{\tenrm Z\kern-.31em{Z}}}}
\def\CcC{{\hbox{\tenrm C\kern-.45em{\vrule height.67em width0.08em depth-
.04em \hskip.45em }}}}

\newcommand{\bc}{\begin{center}}
\newcommand{\ec}{\end{center}}

\begin{document}

\title{Emission of photons through cavity mirrors in the absence of external driving}
\author{A Beige$^1$, A Capolupo$^2$, A Kurcz$^1$,  E Del~Giudice$^3$, and G  Vitiello$\, ^2$}
\address{$^1$The School of Physics and Astronomy, University of Leeds, Leeds, LS2 9JT, United Kingdom}
\address{$^2$Dipartimento di Matematica e Informatica and \\
I.N.F.N., Universit\'a di Salerno,~Fisciano~(SA)-84084,~Italy}
\address{$^3$I.N.F.N. - via Celoria 16 - Milano, Italy}


\begin{abstract}
We present a mechanism of energy concentration in a system composed by an optical cavity and a large number of strongly confined atoms, which cannot be described in the rotating wave approximation. The mechanism consists in the emission of photons through the cavity mirrors even in the absence of external driving.
\end{abstract}

\section{Introduction}
Quantum electrodynamics phenomena in cavities are among the most intriguing topics of quantum optics and quantum information. In particular, the spontaneous emission of photons from a number of atoms trapped inside an optical cavity has been analyzed extensively. Experimental results are well described by models based on the rotating wave approximation (RWA).
However, in some cases the counter-rotating term neglected in the RWA becomes important. It has been shown, for example, that  the  counter-rotating  term relates to the entropy operator and generates an irreversible time evolution \cite{Kurcz:2010wx}.

In this paper, we report on  recent results \cite{Kurcz:2009gh} according to which, under proper conditions,  quantum systems can leak energy into the environment even in  the  absence  of  external  driving. This effect is due to the counter-rotating terms in the interaction Hamiltonian which are ignored in the RWA \cite{Kurcz:2010wx}.

We analyze single and composite quantum systems consisting of many atoms inside an optical resonator. For a large number of atoms inside the cavity, we find that the stationary state photon emission rate can be experimentally detected. Similar phenomena might contribute to the heating in sonoluminescence experiments \cite{SL}.

\section{The model}

We consider the Schr\"odinger picture and the Born and the dipole approximation. The Hamiltonian $H$ describing the evolution of the system
is $H = H_0 + H_{\rm int}$ with
\begin{eqnarray} \label{H}
H_0 &=& \hbar \omega ~s^+ s^- + \sum _{{\bf k},\lambda} \hbar \omega _k ~ a ^\dagger _{{\bf k} \lambda} a _{{\bf k} \lambda}~, \nonumber \\
H_{\rm int} &=& \sum_{{\bf k}, \lambda} \hbar ~ \big( g_{{\bf k} \lambda} ~ a _{{\bf k} \lambda} + \tilde g_{{\bf k} \lambda} ~ a _{{\bf k} \lambda}^\dagger \big) ~ s^+ + {\rm h.c.}
\end{eqnarray}
and $|\tilde g_{{\bf k} \lambda} | = |g_{{\bf k} \lambda}|$. $a_{{\bf k} \lambda}$ are the annihilation operators of photons with polarizations $\lambda$ and frequencies $\omega_k$. In the case of an optical cavity, $\omega \equiv \omega_{\rm c}$ is the frequency of its field mode and $s^\pm$ are the photon creation and annihilation boson operators $c^\dagger$ and $c$.
In the case of a large number of tightly confined two-level atoms with states $|0 \rangle $ and $| 1 \rangle $, $\hbar \omega \equiv \hbar \omega_0$ is the energy of the excited state $|1 \rangle$ of a single atom and $s^\pm$ are related to the collective raising and lowering operators $S^\pm$, with
$\left [ S^-, S^+ \right ] = 1$ and $ \left [ S^+, S^+ \right] = 0 = \left [ S^-, S^- \right]$.

Let us consider the case of atoms confined in a region with linear dimensions much smaller than the wavelength of the emitted light: $|{\bf k} \cdot ( {\bf r}_j - {\bf r}_i )| \ll 1$. Then all particles have approximately the same $g _{{\bf k} \lambda}$ and $\tilde g _{{\bf k} \lambda}$. The structure of the Hamiltonian $H$ remains the same if we replace $s^+s^-$ in $H_0$ by $\sigma_3$ and $s^\pm$ in $H_{\rm int}$ by $\sigma^\pm$ where $\sigma^\pm$ and $\sigma_3$ are given by
\begin{eqnarray} \label{sigma}
\sigma^{\pm} \equiv \sum _{i=1} ^N \sigma _i ^{\pm} ~, ~~
\sigma _3 \equiv \sum_{i=1} ^N \sigma _{3i}
\end{eqnarray}
with $\sigma _{3i} = {1 \over 2} \left ( | 1 \rangle_{ii} \langle 1 |-| 0 \rangle_{ii} \langle 0 | \right )$, $\sigma_i^+ = | 1 \rangle_{ii} \langle 0 |$ and $\sigma_i ^- = | 0 \rangle_{ii} \langle 1 |$ the su(2) spin-like operators of atom $i$:
$
\left [ \sigma _{3i} , \sigma_i ^{\pm} \right ] = \pm \sigma_i ^ {\pm} \, , ~ \left [ \sigma_i ^-  , \sigma_i ^+\right] = - 2 \sigma _{3i} ~ .
$
Assuming that the atoms are initially all in their ground state, under the
action of $\sigma^{\pm} $ and $\sigma _3$, they evolve  into the states:
\begin{eqnarray}
| l \rangle _{\rm p} &\equiv& \left [ | 0_1 0_2 0_3 \dots 0_{N-l} 1_{N-l+1} 1_{N-l+2} \dots 1 \rangle + \dots \right . \left . + | 1_1 1_2 \dots 1_l 0_{l+1} 0_{l+2} \dots 0 \rangle \right ] / \left (
\begin{array}{c} N \\ l \end{array} \right )^{1/2}
\end{eqnarray}
which are the eigenstates of $\sigma_3$. The difference between excited and unexcited particles is counted by $\sigma _3$, since $_{\rm p} \langle l | \sigma _3 | l \rangle _{\rm p} = l - {1 \over 2} N$. For any $l$ we have \cite{cool}:
\begin{eqnarray}\label{Dicke1}
\sigma ^+ ~ | l \rangle _{\rm p} = \sqrt{l+1} \sqrt{N-l}  ~ | l + 1 \rangle _{\rm p} ~ ,  ~~~~
\sigma ^- ~ | l \rangle _{\rm p} = \sqrt{N-(l-1)} \sqrt{l} ~ | l - 1 \rangle _{\rm p} ~ .
\end{eqnarray}
Then $\sigma ^{\pm}$ and $\sigma _3$ are represented on $| l
\rangle_{\rm p}$ by
\cite{Holstein,Shah} $\sigma ^+ = \sqrt{N} S ^+ A_s$, $\sigma ^- = \sqrt{N}
A_s S^- $ with $ \sigma _3 = S^+ S^- - {1 \over 2} N$, $ A_s = \sqrt{1-S^+ S^-
  /N}$, $S^+ | l \rangle _{\rm p} = \sqrt{l+1} | l + 1 \rangle _{\rm p}$, and
$S^- | l \rangle _{\rm p} = \sqrt{l} | l - 1 \rangle _{\rm p}$ for any
$l$. The $\sigma$'s  satisfy the su(2) algebra; however, for
$N \gg l$, Eq.(\ref{Dicke1}) becomes
$
\sigma ^{\pm} ~ |l \rangle _{\rm p} = \sqrt{N} S^{\pm} ~ | l \rangle _{\rm p}
~.$
This implies that for large $N$, the su(2) algebra, in terms of $S^{\pm}$ and $S_3 \equiv \sigma_3$, contracts to the projective algebra e(2) \cite{Shah}
$
\left[ S_3, S^{\pm} \right ] = \pm S^{\pm}  ,  ~ \left [ S^-, S^+ \right ] = 1 ~.
$
The $s^\pm$ operators in the cavity case and in the many atom case are then formally the same.

We assume that the state of a single system $|\varphi \rangle$ at $t=0$ is known. Moreover, we point out that spontaneously emitted photons leave at a very high speed and cannot be reabsorbed. Thus the free radiation field is initially in a state (denoted by $|{\cal O} \rangle$) with only a negligible photon population in the optical regime \cite{Hegerfeldt93,Dalibard,Carmichael}.
In the case of {\em no} photon emission in $(0,\Delta t)$, the (unnormalised) state vector of system and bath is
\begin{equation} \label{kernel}
|{\cal O} \rangle |\varphi^0_{\rm I} \rangle = |{\cal O} \rangle \langle {\cal O}| ~ U_{\rm I} (\Delta t,0) ~ |{\cal O} \rangle |\varphi \rangle\,.
\end{equation}
    By using second order perturbation theory in the interaction picture we find
\begin{equation} \label{kernel2}
|\varphi^0_{\rm I} \rangle = \big[ 1 - A ~ s^+ s^- - B ~ s^- s^+ - C ~ s^{+2} - D ~ s^{-2} \big] ~ |\varphi \rangle
\end{equation}
with
\begin{eqnarray} \label{ABC}
A &=&  \int_0^{\Delta t} \!\! {\rm d}t \int_0^t \!\! {\rm d}t' ~ \sum _{{\bf k}, \lambda} g_{{\bf k} \lambda} \tilde g_{{\bf k} \lambda}^* ~ {\rm e}^{{\rm i}(\omega-\omega_k)(t-t')} ~, \nonumber \\
B &=& \int_0^{\Delta t} \!\! {\rm d}t \int_0^t \!\! {\rm d}t' ~ \sum _{{\bf k}, \lambda} g_{{\bf k} \lambda}^* \tilde g_{{\bf k} \lambda} ~ {\rm e}^{-{\rm i}(\omega+\omega_k)(t-t')} ~, \nonumber \\
C &=& \int_0^{\Delta t} \!\! {\rm d}t \int_0^t \!\! {\rm d}t' ~ \sum _{{\bf k}, \lambda} g_{{\bf k} \lambda} \tilde g_{{\bf k} \lambda} ~ {\rm e}^{{\rm i} (\omega -\omega_k) t + {\rm i} (\omega + \omega_k) t'} ~, \nonumber \\
D &=& \int_0^{\Delta t} \!\! {\rm d}t \int_0^t \!\! {\rm d}t' ~ \sum _{{\bf k}, \lambda} g_{{\bf k} \lambda}^* \tilde g_{{\bf k} \lambda}^* ~ {\rm e}^{- {\rm i} (\omega + \omega_k) t - {\rm i} (\omega - \omega_k) t'}  ~.~~~~~~
\end{eqnarray}
In the case of an emission, the (unnormalised) density matrix of the system  is
\begin{equation} \label{kernel4}
\rho^>_{\rm I} = {\rm Tr}_{\rm R} \left[ \sum _{{\bf k}, \lambda} a ^\dagger _{{\bf k} \lambda} a _{{\bf k} \lambda}
~ U_{\rm I} (\Delta t,0) ~ \tilde \rho ~ U_{\rm I}^\dagger (\Delta t,0)  \right]
\end{equation}
where $\tilde \rho = |{\cal O} \rangle \langle {\cal O}| \otimes \rho$ is the initial state of system and bath. Using  second order perturbation theory we have
\begin{equation} \label{kernel5}
\rho^>_{\rm I} = \tilde A ~ s^- \rho s^+ + \tilde B ~ s^+ \rho s^- + \tilde C ~ s^- \rho s^- + \tilde D ~ s^+ \rho s^+ ~,
\end{equation}
with \cite{Kurcz:2009gh}
\begin{eqnarray} \label{CD}
&& \tilde A = 2 {\rm Re} A ~, ~~ ~~ \tilde C = C^* + {\rm e}^{- 2 {\rm i} \omega \Delta t} ~ C ~,~~~~
   \tilde B = 2 {\rm Re} B ~, ~~ ~~\tilde D = D^* + {\rm e}^{2 {\rm i} \omega \Delta t} ~ D ~,\\
&& C = D^* = {1 \over 2}f ~ \gamma_{\rm C} ~, ~~ ~~ \tilde C = \tilde D^* = f^* ~
\gamma_{\rm C} ~,
~~ ~~ f \equiv  {\rm e}^{{\rm i} \omega \Delta t} ~ \sin(\omega \Delta t) / \omega~,
\end{eqnarray}
and $\gamma_{\rm C}$  real.
The density matrix is obtained by averaging over the subensemble with and the subensemble without photon
emission (cf.~(\ref{kernel2}) and (\ref{kernel5})) at $\Delta t$; we find
\begin{eqnarray} \label{kernel7}
\rho_{\rm I} (\Delta t) &=& \rho - \big[ \big( A ~ s^+ s^- + B ~ s^- s^+) ~ \rho + {\rm h.c.} \big] \nonumber \\
&& \hspace*{-0.4cm} - {1 \over 2} \gamma_{\rm C} ~ \big[ \big( f ~ s^{+2} + {\rm h.c.} \big) ~ \rho + {\rm h.c.} \big] + 2 {\rm Re} A ~ s^- \rho s^+ \nonumber \\
&& \hspace*{-0.4cm} +  2 {\rm Re} B ~  s^+ \rho s^- + \gamma_{\rm C} ~ \big[  f ~ s^+ \rho s^+ + {\rm h.c.} \big] ~.
\end{eqnarray}
By using second order perturbation theory we derive the following master equation:
\begin{eqnarray} \label{deltarho2}
\dot \rho &=& - {{\rm i} \over \hbar} \left[ H_{\rm cond} \rho - \rho H_{\rm cond}^\dagger \right] + {\cal R}(\rho) ~ , \nonumber \\
{\cal R}(\rho) &=& \gamma_{\rm A} ~ s^- \rho s^+ + \gamma_{\rm B} ~ s^+ \rho s^- + \gamma_{\rm C} ~ \big( s^- \rho s^- + {\rm h.c.} \big) ~ , \nonumber \\
H_{\rm cond} &=&- {{\rm i} \over 2} \hbar \big[ \gamma_{\rm A} ~ s^+ s^- + \gamma_{\rm B} ~ s^- s^+ + \gamma_{\rm C} ~ \big( s^{+2} + {\rm h.c.} \big) \big]  +  \hbar {\widetilde \omega} ~ s^+ s^-
\end{eqnarray}
with $\gamma_{\rm A} = 2 {\rm Re} A/\Delta t$, $\gamma_{\rm B} =  2 {\rm Re} B/\Delta t$, and ${\widetilde \omega}$ the shifted bare transition frequency.

\section{Photon emission from a single quantum system} \label{single}

Let us study the possibility of photon emission from a single quantum system. The probability density $I_\gamma =  {\rm Tr} ({\cal R}(\rho))$ for a photon emission of a system prepared in $\rho$ is
\begin{eqnarray} \label{kernel3}
I_\gamma &=& \left \langle \gamma_{\rm A} ~ s^+ s^- + \gamma_{\rm B} ~ s^- s^+ + \gamma_{\rm C} ~ \left( s^{+2} + s^{-2} \right) \right \rangle ~.~~~~
\end{eqnarray}
Such a quantity can be computed by considering the time evolution of the expectation values $\mu_1 \equiv \langle s^+ s^- \rangle$, $\xi_1 \equiv {\rm i} \langle s^{-2} - s^{\dagger +2} \rangle$, and $\xi_2 \equiv \langle s^{-2} + s^{+2} \rangle$. We obtain a closed set of rate equations,
\begin{eqnarray}
&& \dot \mu_1 = - (\gamma_{\rm A} - \gamma_{\rm B}) ~ \mu_1 + \gamma_{\rm B} ~, \nonumber \\
&& \dot \xi_1 = - (\gamma_{\rm A} - \gamma_{\rm B}) ~ \xi_1 + 2 \widetilde \omega ~ \xi_2 ~, \nonumber \\
&& \dot \xi_2 = - (\gamma_{\rm A} - \gamma_{\rm B}) ~ \xi_2 - 2 \tilde \omega ~ \xi_1 - 2 \gamma_{\rm C} ~.
\end{eqnarray}
Setting these derivatives equal to zero, we find that the stationary photon emission rate of a single bosonic system  is
\begin{eqnarray}
I_\gamma = {2 \gamma_{\rm A} \gamma_{\rm B} \over \gamma_{\rm A} - \gamma_{\rm B}} - {2 \gamma_{\rm C}^2 (\gamma_{\rm A} - \gamma_{\rm B}) \over  4 \widetilde \omega^2 + (\gamma_{\rm A} - \gamma_{\rm B})^2} ~.
\end{eqnarray}
$I_\gamma$ is zero in the absence of external driving only when
$
\gamma_{\rm B} = \gamma_{\rm C} = 0 ~.
$
 However, this assumption is based on how the integrals in (\ref{ABC}) are evaluated and whether relations like $\tilde D(\omega) = \tilde C(-\omega)$ are taken into account or not.

\section{Photon emission from a composite quantum system} \label{comp}

Let us now analyze the photon emission from a composite quantum system. We consider a large number $N$ of tightly confined atoms inside an optical cavity. The energy of this composite system in the Born and dipole approximations is $H= H_{0}+ H_{int}$ where
\begin{eqnarray} \label{H2}
H_0 &=& \hbar ~ \omega_{\rm c} ~c^\dagger c + \hbar \omega_0 ~ S^+ S^-  + \sum _{{\bf k},\lambda} \hbar \omega _k ~ a ^\dagger _{{\bf k} \lambda} a _{{\bf k} \lambda} ~ , \nonumber \\
H_{\rm int} &=& \sum_{{\bf k}, \lambda} \hbar \big( g_{{\bf k} \lambda} ~ a _{{\bf k} \lambda} + \tilde g_{{\bf k} \lambda} ~ a _{{\bf k} \lambda}^\dagger \big) ~ c^\dagger + \sqrt{N} \hbar ~ \big( q_{{\bf k} \lambda} ~ a _{{\bf k} \lambda} \nonumber \\
&& \hspace*{-0.5cm} + \tilde q_{{\bf k} \lambda} ~ a _{{\bf k} \lambda}^\dagger \big) ~ S^+ + \sqrt{N} \hbar g_{\rm c} ~ \big( c+c^\dagger \big) ~ S^+ + {\rm h.c.} ~~~~
\end{eqnarray}
with $g_{\rm c}$, $g_{{\bf k} \lambda}$, $\tilde g_{{\bf k} \lambda}$,
$q_{{\bf k} \lambda}$ and $\tilde q_{{\bf k} \lambda}$ being coupling
constants. For simplicity, the cavity photon states should be chosen such that
$g_{\rm c}$ becomes real. Proceeding as in the single system case,
we have again the master equation (\ref{deltarho2}) but with
\begin{eqnarray} \label{last2}
{\cal R}(\rho) &=& \kappa~ c \rho c^\dagger + N\Gamma ~ S^+ \rho S^- ~, \nonumber \\
H_{\rm cond} &=& \hbar \Big( \widetilde \omega_{\rm c} - {{\rm i} \over 2} \kappa \Big) ~ c^\dagger c + \hbar \Big( \widetilde \omega_0 - {{\rm i} \over 2} N \Gamma \Big) ~ S^+ S^- \nonumber \\
&& + \sqrt{N} \hbar g_{\rm c} ~ \big( c+ c^\dagger \big) \big( S^+ + S^- \big)  ~,
\end{eqnarray}
where $\widetilde \omega_{\rm c}$ and $\widetilde \omega_0$ are the bare atom and cavity frequencies, $\kappa$ is the cavity decay rate, and $\Gamma$ is the decay rate of the excited state of a single atom. The crucial difference with respect to the usual Jaynes-Cummings model \cite{Knight} is the presence of the $c S^-$ and the $c^\dagger S^+$ term in (\ref{H2}) which are zero in the RWA. These operators generate a non-zero stationary state population in excited states and the continuous emission of photons, even without external driving.

To calculate this rate, we  neglect $\gamma_{\rm B}$ and $\gamma_{\rm C}$, since this assures that no emissions occur in the absence of external driving in the single system case. We obtain the following closed set of the rate equations:
\begin{eqnarray}
&&\dot \mu _1 = \sqrt{N} g_{\rm c} \eta _1 - \kappa \mu _1 \, , ~~
\dot \mu _2 = \sqrt{N}  g_{\rm c} \eta _2 - N \Gamma \mu _2 ~ , \nonumber \\
&&\dot \eta _1 =  2 \sqrt{N} g_{\rm c} (1 + 2 \mu_2 + \xi_4) + \widetilde \omega_0 \eta _3 + \widetilde \omega_{\rm c} \eta _4 - {\textstyle {1 \over 2}} \zeta \eta_1 ~ , \nonumber \\
&&\dot \eta _2 = 2 \sqrt{N}  g_{\rm c} (1 + 2 \mu_1 + \xi_2) + \widetilde \omega _0 \eta _4 + \widetilde \omega _{\rm c} \eta _3 - {\textstyle {1 \over 2}} \zeta \eta_2 ~ , \nonumber \\
&&\dot \eta _3 = - 2 \sqrt{N}  g_{\rm c} (\xi_1 + \xi_3) - \widetilde \omega _0 \eta _1 - \widetilde \omega _{\rm c} \eta _2 - {\textstyle {1 \over 2}} \zeta \eta _3 ~ , \nonumber \\
&&\dot \eta _4 = - \widetilde \omega _0 \eta _2 - \widetilde \omega_{\rm c} \eta _1 - {\textstyle {1 \over 2}} \zeta \eta_4 ~, \nonumber \\
&&\dot \xi _1 = 2 \sqrt{N} g_{\rm c} \eta_4 + 2 \widetilde \omega_{\rm c} \xi_2 - \kappa \xi_1 ~, \nonumber \\
&&\dot \xi _2 = - 2 \sqrt{N} g_{\rm c} \eta_1 - 2 \widetilde \omega_{\rm c} \xi_1 - \kappa \xi_2 ~, \nonumber \\
&&\dot \xi _3 = 2 \sqrt{N} g_{\rm c} \eta_4 + 2 \widetilde \omega_0 \xi_4 - N \Gamma \xi_3 ~, \nonumber \\
&&\dot \xi _4 = - 2 \sqrt{N} g_{\rm c} \eta_2 - 2 \widetilde \omega_0 \xi_3 - N \Gamma \xi_4
\end{eqnarray}
with $\mu_1 \equiv \langle c^\dagger c \rangle$, $\mu_2 \equiv \langle S^+ S^- \rangle$, $\eta_{1,2} \equiv {\rm i} \langle (S ^- \pm S ^+) (c \mp c^\dagger) \rangle$, $\eta _{3,4} \equiv\langle (S ^- \mp S ^+) (c \mp c^\dagger )\rangle$, $\xi_1 \equiv {\rm i} \langle c^2 - c^{\dagger 2} \rangle$, $\xi_2 \equiv \langle c^2 + c^{\dagger 2} \rangle$, $\xi_3 \equiv {\rm i} \langle S^{-2} - S^{+2} \rangle$, $\xi_4 \equiv \langle S^{-2} + S^{+2} \rangle$, and $\zeta \equiv \kappa + N \Gamma$.
 By using (\ref{kernel3}) and the stationary state of these equations we obtain the cavity photon emission rate
\begin{eqnarray} \label{IN}
I_{\kappa} &=& {N \zeta \kappa g_{\rm c}^2 \left [ \, 8 \zeta g_{\rm c}^2 + \zeta^2 \Gamma + 4 \Gamma \left ( \widetilde \omega _0 - \widetilde \omega_{\rm c} \right )^2 \, \right] \over 16 \zeta^2 g_{\rm c}^2 \widetilde \omega_0 \widetilde \omega_{\rm c} + 2 \zeta^2 \kappa \Gamma \left ( \widetilde \omega _0 ^2 + \widetilde \omega_{\rm c}^2 \right ) + 4 \kappa \Gamma \left( \widetilde \omega _0^2 - \widetilde \omega_{\rm c}^2 \right)^2}
\end{eqnarray}
which applies for $N \Gamma, \sqrt{N} g_{\rm c}, \kappa \ll \widetilde \omega_0,\widetilde \omega_{\rm c}$. Utilizing  the parameters of the recent cavity experiment with $^{85}$Rb \cite{Trupke} and $N=10^4$ we have $I_\kappa =301~\rm{s}^{-1}$ which can be detected experimentally.

\section{Conclusions} \label{conc}

 In conclusion, by analyzing a system composed by an optical cavity and a large number of tightly confined particles without using RWA approximation, we have shown that even un-excited and un-driven quantum systems might constantly leak energy into their environment. For sufficiently many atoms, a detectable signal can be obtained. This effect is due to non-zero decay rates and to the counter-rotating terms in the interaction Hamiltonian which are  neglected in the RWA.

\section*{Acknowledgements}
 This work has been partially supported by the UK Research Council EPSRC, the EU Research and Training Network EMALI, University of Salerno, and INFN.

\medskip

\end{document}